\documentclass[11pt]{article}
\usepackage{erice,epsfig}

\def\beq{\begin{equation}}
\def\eeq{\end{equation}}
\def\bea{\begin{eqnarray}}
\def\eea{\end{eqnarray}}

\def\oms{\omega_1}
\def\oma{\omega_2}
\begin{document}
\vspace*{4cm}
\title{PARAMETRIC GRAVITY WAVE DETECTOR}

\author{\underline{G. GEMME}\,$^1$, A. CHINCARINI\,$^1$, R. PARODI\,$^1$, PH. BERNARD\,$^2$ and E. PICASSO\,$^3$}
\address{$^1$Istituto Nazionale di Fisica Nucleare, via Dodecaneso, 33, I-16146, Genova, Italy\\
$^2$CERN, CH-1211, Geneva 23, Switzerland\\
$^3$Scuola Normale Superiore, piazza dei Cavalieri, 7, I-56126, Pisa, Italy}

\maketitle
\abstracts{Since 1978 superconducting coupled cavities have been proposed as a sensitive detector of gravitational waves.
The interaction of the gravitational wave with the cavity walls, and the resulting motion, induces the transition of some energy
from an initially excited cavity mode to an empty one. The energy transfer is maximum when the frequency of the wave is equal 
to the frequency difference of the two cavity modes.
In 1984 Reece, Reiner and Melissinos built a detector of the type proposed, and used it as a transducer of 
harmonic mechanical motion, achieving a sensitivity to fractional deformations of the order $\delta x/x \approx 10^{-18}$.
In this paper the working principles of the detector are discussed and the last experimental results summarized.
New ideas for the development of a realistic gravitational waves detector are considered; the outline of a possible detector
design and its expected sensitivity are also shown.}

\section{Introduction}
\label{sec:intro}

In this paper we shall discuss the mechanism of the interaction of a gravitational wave 
with a detector based on two coupled electromagnetic cavities. 
In previous works this issue was discussed using the concept of a dielectric tensor associated 
with the gravitational wave.\cite{pr1} The interaction was analyzed in the reference frame where 
the resonator walls were {\em at rest} even in presence of a gravitational perturbation. 
Here we shall analyze the interaction in the proper reference frame attached to the detector and we shall 
therefore consider both the coupling between the wave and the mechanical structure of the detector
and the perturbation induced on the field stored inside the resonator due 
to the time--varying boundary conditions. 

The proposed detector exploits the energy transfer induced by the gravitational wave 
between two levels of an electromagnetic resonator, whose frequencies $\oms$ and $\oma$ are both  
much larger than the characteristic angular frequency $\Omega$ of the g.w. and satisfy the resonance condition 
$\oma-\oms = \Omega$. The interaction between the g.w. and the detector is characterized by a transfer 
of energy and of angular momentum. Since the elicity of the g.w. (i.e. the angular 
momentum along the direction of propagation) is 2, it can induce a transition between the 
two levels provided their angular momenta differ by 2; this can be achieved by putting the 
two cavities at right angle or by a suitable polarization of the electromagnetic field axis 
inside the resonator. In the scheme suggested by Bernard et al. the two levels are obtained 
by coupling two identical high frequency cavities.\cite{bppr1,bppr2} The angular frequency $\oms$ is the 
frequency of the level symmetrical in the fields of the two cavities, and $\oma$ is that of the 
antisymmetrical one. The frequency difference between the symmetric and the 
antisymmetric level is determined by the coupling and can be adjusted by a careful 
resonator design. Since the detector sensitivity is proportional to the square of the resonator 
quality factor, superconducting cavities must be used for maximum sensitivity.

The power transfer between the levels of a resonator made up of two pill-box 
cavities, mounted end-to-end and coupled by a small circular aperture in their common end 
wall, was checked in a series of experiments by Melissinos et al., where the perturbation of 
the resonator volume was induced by a piezoelectric crystal.\cite{rrm1,rrm2} 

Recently the experiment 
was repeated by our group with an improved experimental set-up;\cite{rsi} we obtained an order of 
magnitude sensitivity to fractional deformations of the resonator length as small as $\delta \ell/\ell \approx 
10^{-20}$ Hz$^{-1/2}$.

In this paper we shall discuss the detector's working principles and briefly review the last experimental results obtained 
by our group on the first prototype. Finally a possible detector design, based on two coupled spherical cavities, is discussed
and its expected sensitivity is shown.

\section{Fundamental principles}
\label{sec:fund}

When the resonator's boundary is deformed by an external force the local displacement vector, $\vec u(\vec r,t)$, 
can be expressed as a superposition of the mechanical undamped normal modes $\{\vec \xi_\alpha(\vec r)\}$: 
$\vec u(\vec r,t) = \sum_\alpha q_\alpha(t) \vec \xi_\alpha(\vec r)$.
$q_\alpha(t)$ is the generalized coordinate of the $\alpha_{th}$ mode, obeying the dynamical equation of motion:
\beq
\label{eq:modeeq}
\ddot q_\alpha(t) + \frac{\omega_\alpha}{Q_\alpha}\,\dot q_\alpha(t) + \omega_\alpha^2 q_\alpha(t) = \frac{f_\alpha(t)}{M_\alpha} 
\eeq
and the modes are normalized according to the relation
\begin{equation}
\label{eq:norm}
\int_{Vol} \rho(\vec x) \vec\xi_\alpha(\vec x) \cdot \vec\xi_\beta(\vec x) \, dV = M_\alpha \delta_{\alpha\beta}
\end{equation} 
being $\rho(\vec x)$ the mass density and $M_\alpha$ the reduced mass of the $\alpha^{th}$ mode. For a homogeneous
system $M_\alpha \equiv M$, where $M$ is the mass of the system.

In eq. \ref{eq:modeeq} an empirical damping term, proportional to the velocity, has been included.

$f_\alpha(t)$ is the generalized force, given by
\beq
\label{eq:genfor}
f_\alpha(t) = \int_{Vol} \vec f(\vec r,t) \cdot \vec\xi_\alpha(\vec r) \, dV
\eeq
where $\vec f(\vec r,t)$ is the external force density acting on the system.

For a plane g.w travelling along the $z$ axis the force density, in the proper reference frame attached to the detector, 
has the form:
\begin{equation}
\label{eq:gwfor}
\vec f(\vec x,t) = -\frac{1}{2} \rho(\vec x) \left 
( \ddot h^1_{\,1} x + \ddot h^1_{\,2} y, \ddot h^1_{\,2} x - \ddot h^1_{\,1} y, 0\right )
\end{equation}
where $h^i_j(t)$, is the adimensional amplitude of the wave, and $h^1_{\,1} = - h^2_{\,2}$, $h^1_{\,2} = h^2_{\,1}$.

To study the mechanism of the energy transfer between the two levels of an electromagnetic resonator perturbed 
by a gravitational wave we shall make use of the fact that the electromagnetic field inside the resonator can be 
expanded over the fields of the normalized, orthogonal normal modes $\{\vec E_n(\vec r)\}$ and 
$\{\vec H_n(\vec r)\}$:\,\cite{slater}
\beq
\label{eq:expansion}
\vec E = \sum_{n} \left (\int_V \vec E \cdot \vec E_n \, dV\right ) \vec E_n \; ; \;
\vec H = \sum_{n} \left (\int_V \vec H \cdot \vec H_n \, dV\right ) \vec E_n
\eeq 

For simplicity we shall assume that in the frequency range of interest only two e.m. modes give a significant 
contribution ($n = 1,2$), and that the external force couples strongly only to one mechanical mode ($\alpha = m$).
If we now introduce a perturbation of the resonator boundary and assume that the perturbation is small 
i.e. that we can still expand the fields inside the perturbed volume over the normal modes of 
the unperturbed resonator, we obtain the following set of equations for the magnetic field expansion coefficients:
\beq
\label{eq:fullsys1}
\ddot{\cal H}_1(t) + \frac{\omega_1}{\mathcal{Q}_1} \dot{\cal H}_2(t) + \omega_1^2 {\cal H}_1(t) = 
- \omega_1 q_m(t) \left( C^m_{11}{\cal H}_1(t) + C^m_{12} {\cal H}_2(t) \right )
\eeq
\beq
\label{eq:fullsys2}
\ddot{\cal H}_2(t) + \frac{\omega_2}{\mathcal{Q}_2} \dot{\cal H}_2(t) + \omega_2^2 {\cal H}_2(t) = 
- \omega_2 q_m(t) \left ( C^m_{21} {\cal H}_1(t) + C^m_{22}{\cal H}_2(t) \right )  
\eeq
\beq
\label{eq:fullsys5}
\ddot q_m(t) + \frac{\omega_m}{Q_m}\,\dot q_m(t) + \omega_m^2 q_m(t) = \frac{f_{m}(t)}{M} + \frac{f_{m}^{ba}(t)}{M} 
\eeq
where we have defined the time--dependent expansion coefficients as ${\cal E}_i(t) \equiv 
\sqrt{\epsilon_0} \int_V \vec E \cdot \vec E_i \, dV$ ,	
${\cal H}_i(t) \equiv \sqrt{\mu_0} \int_V \vec H \cdot \vec H_i \, dV$ and the coupling coefficients as:
\beq
\label{eq:gammap}
C^{m}_{ij} \equiv \int_{{S}} (\omega_i\vec {H}_i \cdot \vec {H}_j - \omega_j\vec {E}_i \cdot 
\vec {E}_j) \, \vec \xi_{m} \cdot d\vec S
\eeq
The integral in eq. \ref{eq:gammap} is made on the {\em unperturbed}
surface of the resonator.

The electromagnetic quality factor ${\cal Q}_n=G_n/R_s$, takes into account the dissipation arising from 
the finite conductivity of the walls.
$R_s$ is the material--dependent surface resistance of the walls, and the geometric factor $G_n$ of the 
$n_{th}$ e.m. mode is given by:
\beq
\label{eq:gdef}
G_n = \omega_n \mu_0 \frac{\int_V H_n^2 \, dV}{\int_S H_n^2 \, dS} 
\eeq
In the following we shall assume $\mathcal{Q}_1 \approx \mathcal{Q}_2 \equiv \mathcal{Q}$.

The term $f_m^{ba}$, in the r.h.s. of eq. \ref{eq:fullsys5}, 
describes the deformation of the walls induced by the stored e.m. fields,
i.e. a back--action effect of the e.m. field on the detector's boundary:
it is well known that in a resonant cavity the stored magnetic field interacts with the rf wall current, 
resulting in a Lorenz force which causes a deformation of the cavity shape.\cite{slater,padam} The radiation pressure is given by:
\beq
\label{eq:radp} 
P_L(\vec r,t) = \frac{1}{2}\left ( \mu_0 \vec H(\vec r,t)^2 - \epsilon_0 \vec E(\vec r,t)^2 \right )
\eeq
Expanding again the fields $\vec E$ and $\vec H$ in terms of the normal modes (eq. \ref{eq:expansion}) we get:
\bea
\label{eq:radpnew} 
\lefteqn{P_L(\vec r,t) =} \nonumber \\
& &  \frac{1}{2}\left ( \mathcal{H}_1(t)^2 \vec H_1(\vec r)^2 + \mathcal{H}_2(t)^2 \vec H_2(\vec r)^2  +
2 \, \mathcal{H}_1(t) \mathcal{H}_2(t) \vec H_1(\vec r) \cdot \vec H_2(\vec r) \right ) -  \nonumber \\
& & \frac{1}{2}\left ( \mathcal{E}_1(t)^2 \vec E_1(\vec r)^2 + \mathcal{E}_2(t)^2 \vec E_2(\vec r)^2  +
2 \,\mathcal{E}_1(t) \mathcal{E}_2(t) \vec E_1(\vec r) \cdot \vec E_2(\vec r) \right )
\eea

Since $\mathcal{H}_1(t) \sim \exp(j\omega_1 t)$, $\mathcal{H}_2(t) \sim \exp(j\omega_2 t)$, 
$\mathcal{E}_1(t) \sim \exp(j\omega_1 t)$ and $\mathcal{E}_2(t) \sim \exp(j\omega_2 t)$ 
only the cross-product terms will give a significant contribution at the resonance frequency $\Omega = \omega_2 - \omega_1$.
The other, rapidly oscillating terms, will just give an average deformation of the detector's walls, 
determining a static frequency shift of the resonant modes, which can easily be compensated 
by an external tuning device.\cite{padam}

The generalized back--action force (cfr. eqs. \ref{eq:genfor} and \ref{eq:fullsys5}), acting on the $m_{th}$ 
mechanical mode of the structure, will be given by:
\bea
\label{eq:bafor}
\lefteqn{f^{ba}_m =}\nonumber \\
& & \frac{1}{2} \mathcal{H}_1(t) \mathcal{H}_2(t) \int_{S} \left(\vec H_1 \cdot \vec H_2\right) \, \vec \xi_m \cdot d\vec S
- \frac{1}{2} \mathcal{E}_1(t) \mathcal{E}_2(t) \int_{S} \left(\vec E_1 \cdot \vec E_2\right) \, \vec \xi_m \cdot d\vec S 
\equiv \nonumber \\
& & \frac{1}{2}\mathcal{A}^{m}_{21} \, \mathcal{H}_2(t) \mathcal{H}_1(t) - 
\frac{1}{2}\mathcal{B}^{m}_{21} \,\mathcal{E}_2(t) \mathcal{E}_1(t) 
\eea
where we have introduced the coefficients $\mathcal{A}^{m}_{21}\equiv\int_{S} \left(\vec H_1 \cdot \vec H_2\right)\, \vec \xi_m \cdot d\vec S$ and $\mathcal{B}^{m}_{21}\equiv\int_{S} \left(\vec E_1 \cdot \vec E_2\right)\, \vec \xi_m \cdot d\vec S$.

The analysis of the system of differential equations \ref{eq:fullsys1}--\ref{eq:fullsys5} can be simplified
if we neglect the small perturbation on the initially excited e.m. mode (say mode 1), just taking into account the effects
on the initially empty mode. Furthermore we shall consider the coupling between two TE modes of a resonator: 
for these modes we have vanishing electric field on the resonator surface and, as can readily be calculated, 
$C^{m}_{22} = 0$ and $\mathcal{A}^{m}_{21} \approx C^{m}_{21}/\omega_2$.

With this assumptions we can recast the coupled system of equations in the following form:
\beq
\label{eq:simsys1}
\ddot{\cal H}_2(t) + \frac{\omega_2}{\mathcal{Q}} \dot{\cal H}_2(t) + \omega_2^2 {\cal H}_2(t)= 
-\omega_2 q_m(t)C^m_{21} {\cal H}_1(t)  
\eeq
\beq
\label{eq:simsys2}
\ddot q_m(t) + \frac{\omega_m}{Q_m}\,\dot q_m(t) + \omega_m^2 q_m(t) = \frac{f_{m}(t)}{M} +
\frac{1}{2}\frac{C^{m}_{21}}{M\omega_2} \, {\cal H}_2(t) {\cal H}_1(t)
\eeq

The solution of quations \ref{eq:simsys1}--\ref{eq:simsys2} is straightforward
when the back--action term is switched off in eq. \ref{eq:simsys2} and if we assume
${\cal H}_1(t) = \Re{(A\exp(i\omega_1 t))}$. In this case we have the following 
asymptotic solution (in the frequency domain):
\beq
\label{eq:soluq1}
q_m(\omega) = \frac{f_m(\omega)/M}{(-\omega^2+\omega_m^2+i\,\omega\omega_m/Q_m)}
\eeq 
and 
\bea
\label{eq:soluh1}
\lefteqn{{\cal H}_2(\omega) =}\nonumber \\
& & -\frac{1}{2}\frac{\omega_2 A C^m_{21}}{-\omega^2 + \omega_2^2 +i\,\omega\omega_2/{\cal Q}} \,
\frac{f_m(\omega - \omega_1)/M}{-(\omega-\omega_1)^2 +\omega_m^2 +i\,\omega_m(\omega-\omega_1)/{Q_m}} + \nonumber \\
& & -\frac{1}{2}\frac{\omega_2 A C^m_{21}}{-\omega^2 + \omega_2^2 +i\,\omega\omega_2/{\cal Q}} \,
\frac{f_m(\omega + \omega_1)/M}{-(\omega+\omega_1)^2 +\omega_m^2 +i\,\omega_m(\omega+\omega_1)/{Q_m}}
\eea 

The field amplitude will be maximum when $\omega \approx \oma$. If we design our detector so that $\oma -\oms =\omega_m $ we have:
\beq
\label{eq:soluh2}
\mathcal{H}_2(\omega_2) \approx \frac{1}{2} A C^m_{21}\frac{f_m(\omega_m)}{M}\frac{\mathcal{Q}Q_m}{\oma\omega_m^2}
\eeq 

Eq. \ref{eq:soluh2} shows that the amplitude of field in the initially empty mode is proportional to the amplitude of the field in the excited mode $A$, and to the electromagnetic ($\mathcal{Q}$) and mechanic ($Q_m$) quality factor of the system.

A detailed dicussion of the derivation of equations \ref{eq:fullsys1}--\ref{eq:fullsys5} and of the noise sources 
affecting the performance of the detector will be given elsewhere.

\section{Experimental results}
\label{sec:exp}

\begin{figure}[hbt]
\begin{center} \mbox{\epsfig{file=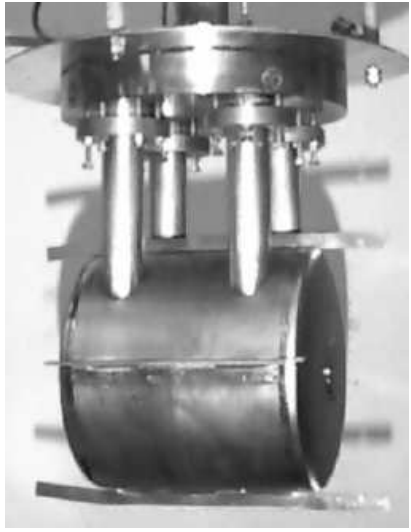,height=7cm}} \end{center}
\caption{Superconducting cavity mounted on the test cryostat.} 
\label{fig:paco1} 
\end{figure} 

The electromagnetic properties of a prototype detector, made up of two pill--box cavities, mounted end--to--end,
and coupled trough an iris on the axis, were measured in a vertical cryostat after 
careful tuning of the two cells frequencies. In order to get maximum sensitivity we need to have two {\it identical}
coupled e.m. resonators, or, in other words, a flat field distribution between the two cavities. 
The symmetric mode frequency was measured at 3.03 GHz and the mode separation was 1.38 MHz.

In order to suppress the noise coming from the symmetric mode at the detection frequency, the transmission detection scheme, with
two magic--tees, was used, as described elsewhere.\cite{rsi}

\begin{figure}[hbt]
\begin{center} \mbox{\epsfig{file=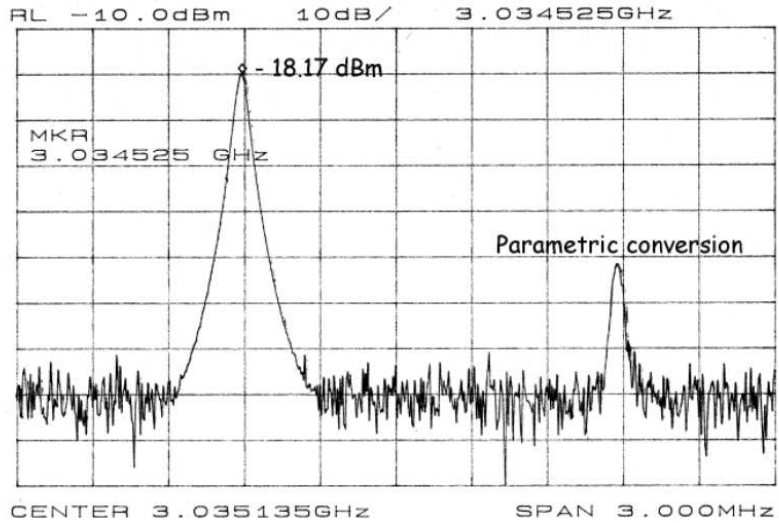,height=7cm}} \end{center}
\caption{Transmission of the symmetric mode (no optimization) measured at the $\Delta$ port of the output magic--tee.} 
\label{mis1} 
\end{figure}

In figure \ref{mis1} the signal from the $\Delta$ port of the output magic--tee is shown for an input power 
$P_i = 1$ W and no adjustments made on the phase and amplitudes of the rf signal entering and leaving the 
cavity. The overall attenuation of the symmetric mode is $R \approx -48$ dB. 

\begin{figure}[hbt]
\begin{center} \mbox{\epsfig{file=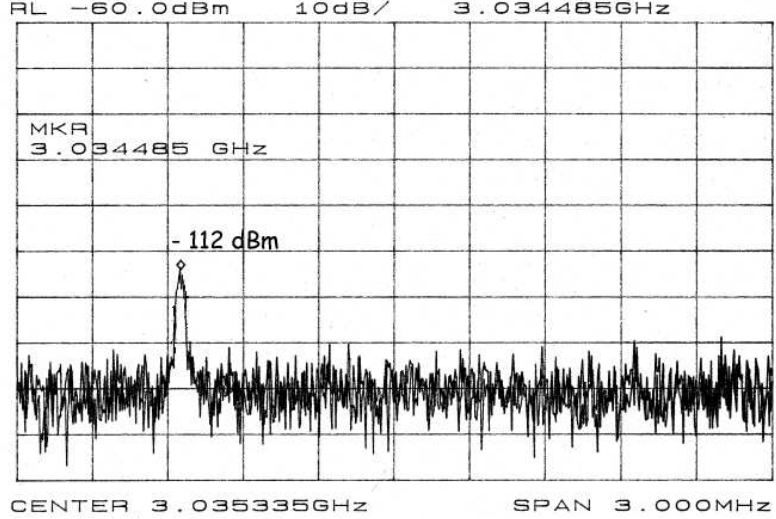,height=7cm}} \end{center}
\caption{Transmission of the symmetric mode in the optimized system, measured at the $\Delta$ 
port of the output magic--tee. Measurement taken with 1 kHz resolution bandwidth.} 
\label{mis3} 
\end{figure}

After balancing the arms of the two magic--tees in order to launch the 
symmetrical mode at the cavity input and to pick up the antisymmetrical one 
at the cavity output, with 1 W (30 dBm) of power at the $\Sigma $ port of the 
first magic--tee, $6.3 \times 10^{-15}$ W (-112 dBm) were detected at the $\Delta$ port of the second 
one, giving an overall attenuation of the symmetric mode of $R \approx -140$ dB (see figure \ref{mis3}).
At a detection frequency of $\Omega / 2 \pi \approx 1$ MHz the sensitivity of the system is quite independent 
from the value of $R$, because of the high cavity $\cal{Q}$. 
Nevertheless for lower frequencies, in a range $\Omega \leq 10$ kHz, where astrophysical sources of 
gravitational waves are expected to exist, this noise source can become dominant and the achieved 
rejection is fundamental in order to pursue the design of a working g.w. detector in the 1--10 kHz frequency range.    

The cavity loaded quality factor was $\mathcal{Q}_L = 1 \times 10^9$ at 1.8 K,
and the energy stored in the cavity with 10 W input power was approximately 1.8 J 
(limited by the maximum power delivered by the rf amplifier), 
with both the input and output ports critically coupled ($\beta_1 \approx \beta_2 \approx 1$).

To excite the antisymmetric mode a piezoelectric crystal (Physik Instrument PIC 140, with longitudinal piezoelectric 
coefficient $\kappa_\ell = 2 \times 10^{-10}$ m/V) was fixed to one cavity wall. The driving signal to the crystal 
was provided by a synthesized oscillator with a power output in the range 2--20 mW (3--13 dBm). The oscillator output 
was further attenuated to reduce the voltage applied to the piezo by a series of fixed attenuators and a variable attenuator 
(10 dB step). The oscillator frequency was carefully tuned to maximize the energy transfer between the cavity modes.

The signal emerging from the $\Delta $ port of the output magic--tee was 
amplified by the LNA (48 dB gain) and fed into a spectrum analyzer. In figure \ref{mis1} an example of the parametric conversion 
process is shown. 

The minimum detected noise signal level at the antisymmetric mode frequency, with no excitation coming from the piezo, 
was $P_{out}(\oma) = 5 \times 10^{-19}$ W in a bandwidth $\delta f = 100$ Hz, giving a noise power spectral density 
${\mathcal P}_{out}(\oma) = 5 \times 10^{-21}$ W/Hz; the main contribution to this signal was the johnson noise of 
the rf amplifier used to amplify the signal picked from the $\Delta$ port of the output magic--tee.

Taking into account the input and output coupling coefficients the sensitivity if the system is given by
$h_{min} \approx 3 \times 10^{ -20} \mathrm{(Hz)}^{-1/2}$

\section{Future perspectives}
\label{sec:future}

\begin{figure}[hbt]
\begin{center} \mbox{\epsfig{file=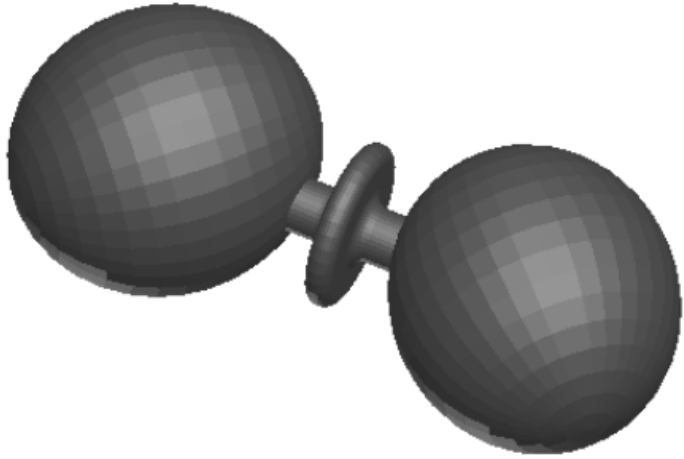,height=7cm}} \end{center}
\caption{Artistic view of the coupled spherical cavities with the central tuning cell.} 
\label{fig:paco2} 
\end{figure} 

The second phase of the R\&D program is focused on the development of a 
detector based on two spherical coupled cavities (see figure \ref{fig:paco2}). In order to approach the 
interesting frequency range for g.w. detection, the mode splitting (i.e. the detection 
frequency) will be $\oma - \oms \approx 10$ kHz. The internal radius of the spherical cavity will be $r \approx 
100$ mm, corresponding to a frequency of the TE$_{011}$ mode $\omega \approx 2$ GHz. The overall system mass and length
will be $M \approx 5$ kg and $L \approx 0.8$ m. The choice of these frequencies for the resonator and mode splitting
will be also useful in order to test the feasibility of a detector working at $\approx 200$ MHz and at a detection frequency
of $\approx 1$ KHz. 

A tuning cell, or a 
superconducting bellow, will be inserted in the coupling tube between the two cavities, 
allowing to tune the coupling strength (i.e. the detection frequency) in a narrow range 
around the design value. 


The choice of spherical cells depends on several factors:

\begin{itemize}
\item{From the point of view of the electromagnetic design the spherical cell has the 
highest geometrical factor, and so the highest quality factor, for a given surface 
resistance.}
\end{itemize}

For the TE$_{011}$ mode of a sphere the geometric factor $G$ has a value $G \approx 850 \, \Omega$, while 
for a standard elliptical accelerating cavity the TM$_{010}$ mode has a value of $G \approx 250 \, \Omega$. 
Looking at the best reported values of quality factor of accelerating cavities, which 
typically are in the range $10^{10}-–10^{11}$, we can extrapolate that the quality factor of the TE$_{011}$ 
mode of a spherical cavity can exceed $\mathcal{Q} \approx 10^{11}$.

\begin{itemize}
\item{From the mechanical point of view it is well know that a sphere has the highest 
interaction cross-section with a g.w. and that only a few mechanical modes of the 
sphere do interact with a gravitational perturbation (the quadrupolar ones).\cite{lobo}}
\end{itemize}

\begin{figure}[hbt]
\begin{center} \mbox{\epsfig{file=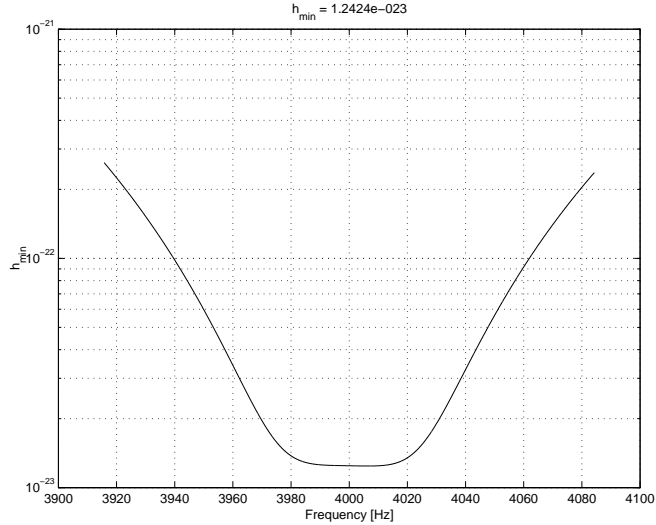,height=7cm}} \end{center}
\caption{Calculated system sensitivity for a periodic source after 1 year integration time
($\omega_m = \oma - \oms = 4$ kHz, $\mathcal{Q}=10^{11}$, $Q_m=5\times 10^{3}$, $T=1.8$ K).} 
\label{fig:hmin4} 
\end{figure}

\begin{figure}[hbt]
\begin{center} \mbox{\epsfig{file=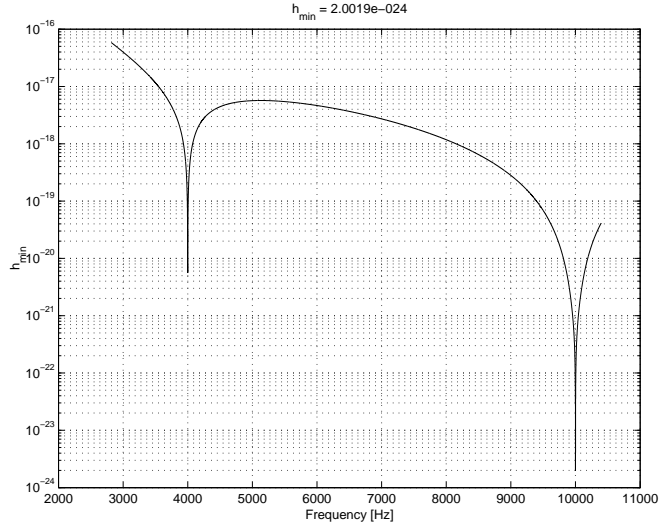,height=7cm}} \end{center}
\caption{Calculated system sensitivity for a periodic source after 1 year integration time
($\omega_m = 4$ kHz, $\oma - \oms=10$ kHz, $\mathcal{Q}=10^{11}$, $Q_m=5\times 10^{3}$, $T=1.8$ K).} 
\label{fig:hmin410} 
\end{figure}

The mechanical design is highly simplified if the spherical geometry is used since the 
deformation of the sphere is given by the superposition of just one or two normal modes of 
vibration and thus can be easily modeled. In fact the proposed detector acts essentially as a 
standard g.w. resonant bar detector: the gravitational perturbation interacts with the 
mechanical structure of the resonator, deforming it. The e.m. field stored inside the 
resonator is affected by the time--varying boundary conditions and a small quantity of 
energy is transferred from the initially excited e.m. mode to the initially empty one, 
provided the g.w. frequency equals the frequency difference of the two modes. A possible 
design of the detector makes use of both the mechanical resonance of the resonator 
structure, and the e.m. resonance. This can be accomplished if the detector is designed in 
order to have the mechanical mode frequency equal to the e.m. modes 
frequency difference $\omega_m = \oma - \oms$. In particular, for the detector designed to work
in the 10 kHz frequency range, the two lowest quadrupolar modes frequency will be 
approximately at 4 and 17 kHz.
The expected sensitivities of the detector for $\oma - \oms = 4$ kHz and $\oma - \oms = 10$ kHz 
are shown in figures \ref{fig:hmin4} and \ref{fig:hmin410}.
In the calculation of the above curves the brownian motion contribution to the detector noise as well
as the noise coming from the detection electronics has been taken into account. Note that, also when
$\oma - \oms \neq \omega_m$ (fig. \ref{fig:hmin410}) the sensitivity of the system is fairly good.
\footnote{Actually the sensitivity of the system at 10 KHz is {\em better} than the sensitivity at 4 KHz.
This is essentially due to the lower value of the brownian noise at higher frequency.}
 
\begin{itemize}
\item{The spherical cells can be esily deformed in order to remove the unwanted e.m. 
modes degeneracy and to induce the field polarization suitable for g.w. detection.}
\end{itemize}

The interaction between the stored e.m. field and the time-varying boundary 
conditions is not trivial and depends both on how the boundary is deformed by the external 
perturbation and on the spatial distribution of the fields inside the resonator. It has been 
calculated that the optimal field spatial distribution is with the field axis of the two cavities 
orthogonal to each other. Different spatial distributions (e.g. with the field axis along the 
resonators' axis) give a smaller effect or no effect at all.


\begin{itemize}
\item{The spherical shape can be easily used to investigate whether the niobium-on-
copper technique could be useful for the detector final design.}
\end{itemize}

The choice between bulk niobium or niobium--on--copper for the final detector design 
has not yet been made and is still under investigation. Both techniques present in principle 
advantages and drawbacks. A prototype of two coupled spherical cavities in bulk niobium 
will be built at CERN in 2002. A single cell, seamless, copper spherical cavity has been 
built at INFN-LNL by E. Palmieri and will be sputter coated at CERN.

\section{Conclusions}
\label{sec:concl}

A first prototype of the detector,  made up of two pill-box cavities, mounted end-to-
end, has been built and successfully tested.
A detector based on two coupled spherical cavities is now being designed, and 
preliminar tests on nomal conducting prototypes are being made. The planned timeline is 
as follows:

\begin{itemize}
\item{In 2002 a bulk niobium detector (two spherical cavities, $\omega = 2$ GHz, $\Omega = 10$ kHz, 
fixed coupling) will be built at CERN;}
\item{In 2003 a variable coupling detector will be built and tested.}
\end{itemize}

If experimental results will be encouraging, by the end of 2003 a proposal for the 
realization of a g.w. detector, based on superconducting rf cavities will be made.

\section*{Acknowledgements}

Several people gave a significant contribution to this work. In particular we wish to thank
Prof. C.M. Becchi for his useful suggestions and Prof. A.C. Melissinos for his constant interest
in our work and for the fruitful discussions that took place in Erice.

\section*{References}
\bibliographystyle{unsrt}
\bibliography{paco}

\end{document}